**What We Could Lose Trying to Make Mars Habitable**

*The rock, ice, and atmosphere of Mars hold a still largely unread archive. Terraforming could erase it before the biggest scientific questions are answered.*

Ari Koeppel, Lonneke Roelofs, Santiago Vrech, Alicia M. Rutledge, Martin Turbet

Carl Sagan once said that if he could travel back in time, he would visit the Library of Alexandria, where "began the intellectual adventure which has led us into space" [*Sagan et al.*, 1980]. Its lost volumes recorded centuries of history and observation that will never be replaced.

The planet Mars is also a sort of cosmic library, one now under threat. Written across its rocks, ice, and atmosphere is a unique record of 4.5 billion years of solar system history that scientists have only just began to decipher. Six decades of missions have revealed possible evidence of ancient oceans, rivers, snow, and lava, as well as complex organic molecules [e.g., *Murphy et al.*, 2026]. Yet we still do not know why exactly Mars became the cold and dry world it is today while Earth stayed temperate or whether it preserves definitive evidence of life. Terraforming, or the transformation of Mars into a more Earth-like habitat, could erase the remaining clues to those questions forever.

A recent wave of studies backed with private funding suggests a new technique that might, for the first time, be technologically within reach. The studies suggest that releasing metallic nanoparticles into Mars's atmosphere could raise surface temperatures by at least 30°C within a century [e.g., *DeBenedictis et al.*, 2025; *Richardson et al.*, 2026]. The new terraforming approach is distinct from others whose feasibility has been debunked, for example, liberating Mars's own sequestered carbon dioxide ($CO_2$) reservoirs to simulate the greenhouse effect. While there is simply not enough water and $CO_2$ left on Mars to generate any substantial warming [*Jakosky and Edwards*, 2018], the newer proposal would import an atmospheric insulator.

Meanwhile, the heavy-lift rockets needed to deploy these glitter bombs at reduced cost are maturing quickly, as the now publicly financed SpaceX continues to develop its Starship system explicitly to enable Mars settlement. For the first time, engineering a second habitable world looks less like science fiction and more like a project someone alive today might actually attempt. As such, now is the time to ask whether this is a road humanity should actually go down anytime soon. Doing so before the scientific community has thoroughly investigated Mars's archive would be like burning the Library of Alexandria after opening only a single book.

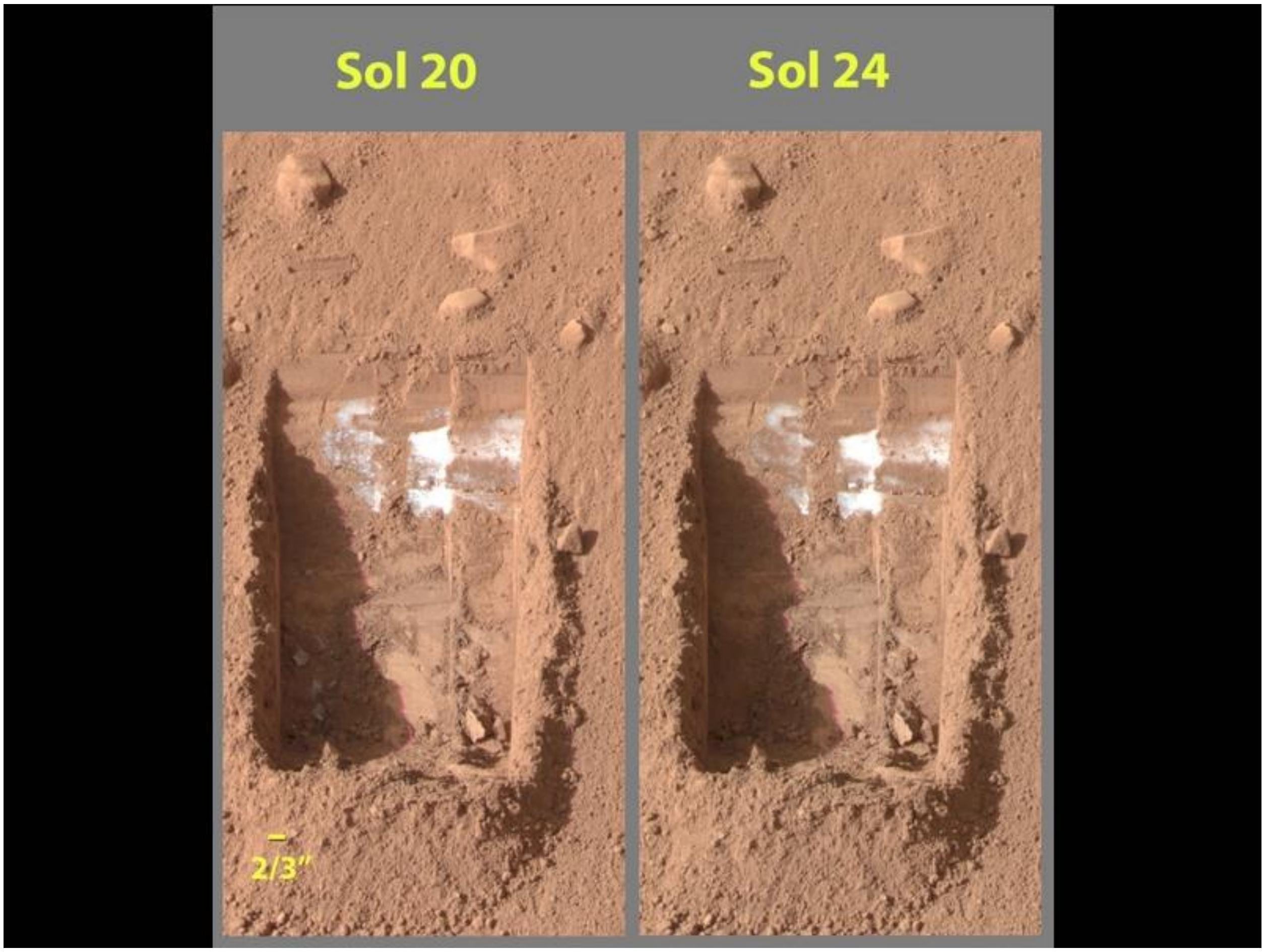

**Figure 1.** In 2008, NASA's Phoenix Lander made the only direct observations of water ice from the surface of Mars with these images. Within a few days, the exposed ice completely sublimated. Credit: NASA/JPL-Caltech/University of Arizona/Texas A&M University https://science.nasa.gov/resource/ice-on-mars-now-its-gone/

## The Martian Library at Risk

The most vulnerable volumes of the Martian library are the climate records preserved in ice. On Earth, ice cores from Greenland and Antarctica revolutionized climate science by revealing how the atmosphere and temperature changed together through time. Cores of Martian ice have the same revolutionary potential, but no mission thus far has sampled them.

The Martian polar layered deposits encode up to 30 million years of climate history in precisely layered strata of ice and dust, likely the longest accessible ice record in our solar system [*Phillips et al.*, 2008; *Becerra et al.*, 2019]. Buried ice in the planet's middle latitudes may extend the record by another 470 million years [*Arnold and Butcher*, 2025].

That archive likely preserves a detailed isotopic catalog of how Mars's climate yo-yoed as a result of changes in its orbit and axial tilt; how dust, $CO_2$, and water have cycled between the atmosphere and surface; and whether conditions were ever suitable for life. Reading that record would tell us not only about Mars but also about how planetary systems generally respond to large perturbations, with immediate relevance to our now rapidly changing Earth.

Water ice is thought to lie at the surface or beneath sediment across roughly half of Mars—about 70 million square kilometers [*Morgan et al.*, 2025]. Sampling the diversity of deposits contained within the vast expanse of underground ice will take time, and even a few early samples are likely to generate more questions than they answer. Interpretation of the record depends on the relationships between layers, not one isolated core. Rapid planetwide warming would irreversibly disturb that stratigraphic context.

What's more, the ice may contain the most tantalizing target of all: evidence of active or past life. If microbial life has ever existed on Mars, ice and the salty liquid along its fringes may be its most promising habitat [*Orosei et al.*, 2018]. Confirming that life is not unique to Earth would rank among the most significant discoveries in the history of science.

Warming Mars would destabilize this entire archive and potential microbial habitat. In a matter of a few years, ice would melt or sublimate, surface minerals untouched by liquid water for billions of years would be irreversibly altered, preserved complex molecules would undergo rapid reactions, and the isotopic clues in the atmosphere would spoil.

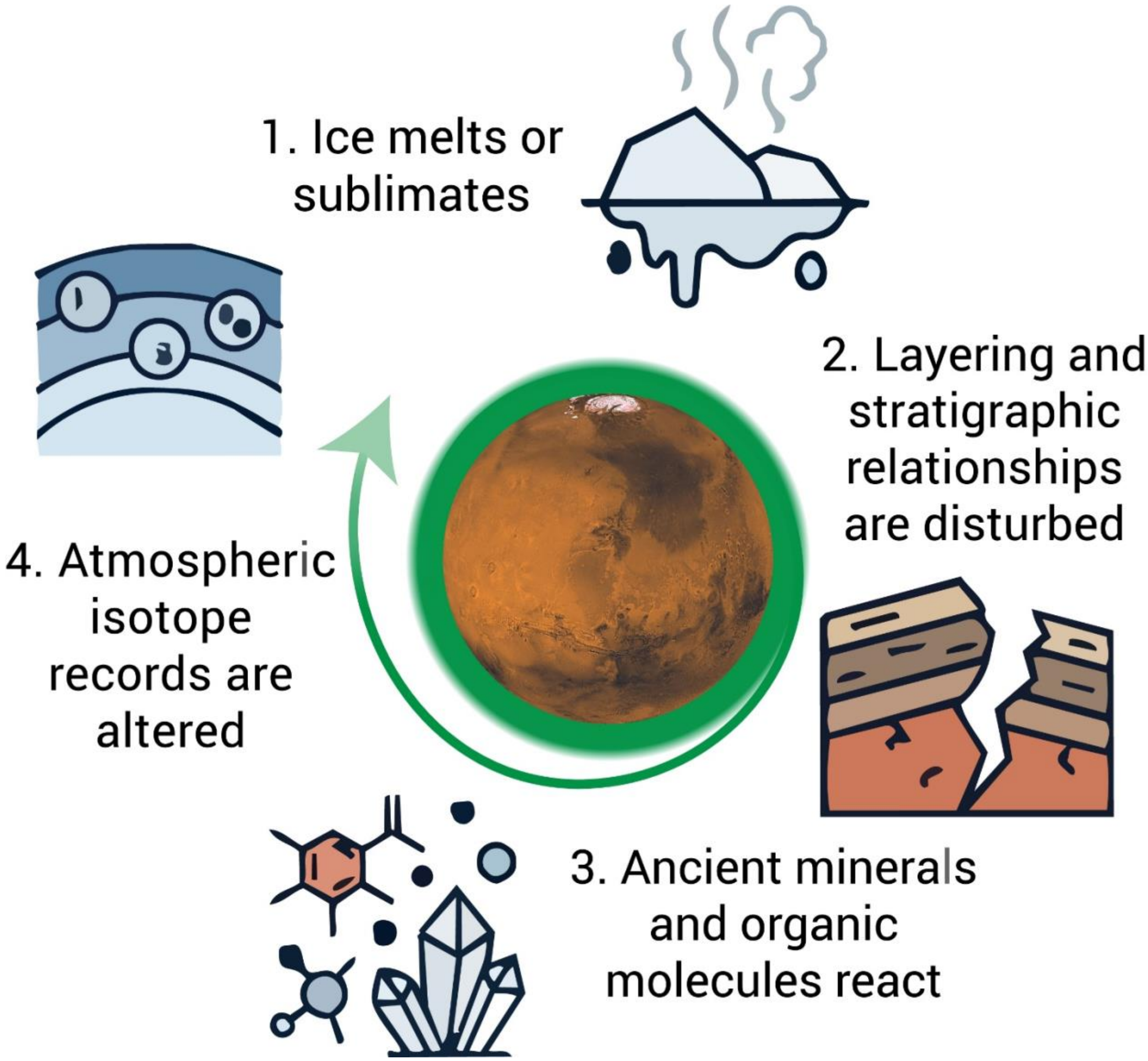

**Figure 2.** Once Mars terraforming is initiated, a series of irreversible consequences will follow. Credit: Ari Koeppel, with Mars image by NASA/JPL/USGS https://www.jpl.nasa.gov/images/pia00407-global-color-views-of-mars/

### Who Speaks for Mars?

Some of the motivation behind terraforming research seems to come from an ethical imperative to extend the longevity of our species—in this case by securing a second home for humanity as a backstop against global catastrophe. Beyond the technical limitations [*Forget et al.*, 2024], a major dilemma lies in determining who will oversee and who will benefit from this endeavor and whether any given action is worth its costs to both people alive today and potential future knowledge.

For decades, Mars exploration has been guided by community consensus. NASA's plan, Expanding the Horizons of Mars Science (2024–2044), distills input from hundreds of scientists into four driving goals: determining whether Mars ever supported life, understanding its climate history, understanding its evolution as a geological system, and preparing for human exploration. Other space agencies have adopted similar priorities. These goals center around knowledge, and they were created through relatively open debate within the science and exploration community, as well as in discussion with the government and general public.

Privately funded terraforming studies are not beholden to those guidelines. They might proceed with different goals, guardrails, and standards for transparency and accountability, driven by the zeal of a few powerful decisionmakers.

Private funding is not inherently the problem; it has enabled pivotal Earth and space science advances, such as the discovery of galaxies beyond the Milky Way and a more detailed accounting of the global carbon cycle. But unless the funded projects pay careful attention to the perspectives of the broader research community, these funding streams can allow individual ambitions to advance without the technical review, ethical oversight, and public accountability that are typically integrated into agency programs.

**Two Steps to Protect the Martian Library**

Regulation can help ensure that public and private planetary activities are aligned in their mandates. Stakeholders can build upon the infrastructure laid out by the Committee on Space Research (COSPAR) Policy on Planetary Protection.

COSPAR's multilateral planetary protection framework defines how spacefaring nations can best comply with the 1967 Outer Space Treaty's Article IX, which states that signatories should "conduct exploration of [celestial bodies] so as to avoid their harmful contamination." The policy is regularly updated to match the latest scientific and ethical understanding of the issue by consensus of the internationally representative Panel on Planetary Protection (PPP).

However, COSPAR's existing framework falls short of addressing the potential realities of a global-scale terraforming effort, creating a planetary protection gap. PPP has thus far included detailed recommendations against *biological* contamination of the most likely host worlds for extraterrestrial life—Mars, Europa, and Enceladus—but does not explicitly address deliberate, planetary-scale environmental modification to those worlds, which can be considered another form of contamination with substantial consequences for local biology and scientific exploration.

Moreover, the Outer Space Treaty also provides a stopgap against any activity or experiment that "would cause potentially harmful interference with activities in the peaceful exploration and use of outer space." Planetary-scale modification carried out without consensus could plausibly raise concerns under the definition of interference in others' ability to peacefully explore that world.

In August 2026, scientists at the 46th Scientific Assembly of COSPAR in Florence, Italy, began discussion over more clearly delineating what "harmful contamination" in the context of Mars. It would be timely for the PPP to convene a working group, drawing from planetary scientists, ethicists, and legal scholars, to draft a policy covering the alteration of planetary oceans and atmospheres and destruction of icy and geological records. National regulators would then have the scaffolding for enforcing planetary alteration mitigation in mission approval and launch licensing, consistent with the Outer Space Treaty's requirement that governments authorize and supervise the private space activities of companies within their jurisdiction.

The next step is to fund the science before irreversible decisions are made. Mission concepts that could read the icy archive already exist, including the International Mars Ice Mapper Mission (an ice-penetrating radar orbiter), Mars Life Explorer (with a first-of-its-kind deep-drilling tool), and sample return architectures similar to NASA's Mars Sample Return campaign (which was canceled in January 2026). Space agencies should elevate the polar and ice science objectives already identified in their community plans now while pristine access is still possible. Robotic missions first, followed by targeted human expeditions, could translate the ice record into a readable history.

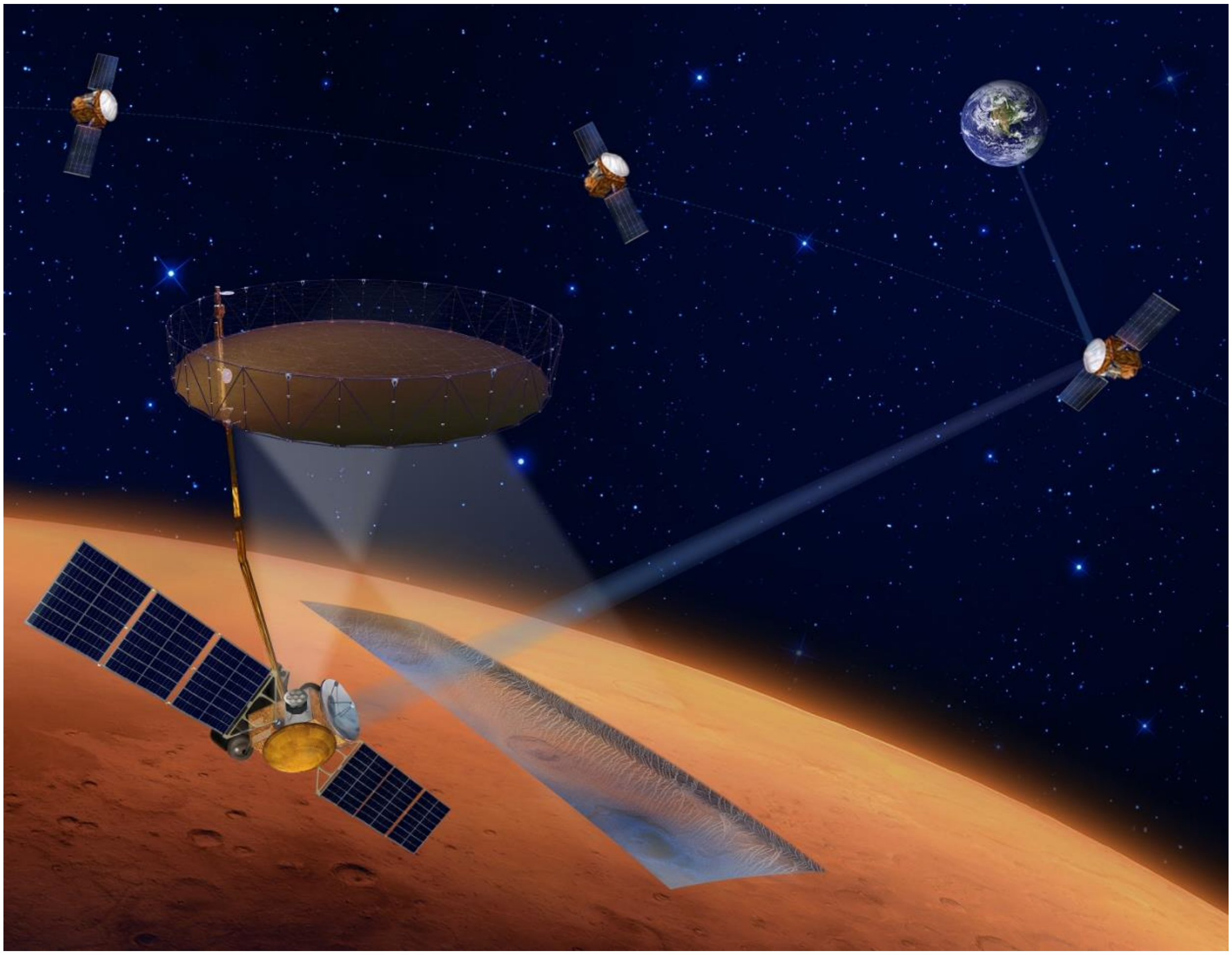

**Figure 3.** An artist's concept shows the proposed International Mars Ice Mapper spacecraft surveying Mars. The mission concept would use orbital radar to map the location, depth, and structure of near-surface water ice, helping scientists reconstruct the planet's climate history and identify sites for further investigation. Credit: NASA https://www.nasa.gov/solar-system/nasa-international-partners-assess-mission-to-map-ice-on-mars-guide-science-priorities/

**Sagan's Charge**

Historians still debate exactly how the Library of Alexandria was lost, but one lesson survives every version of the story: Knowledge was treated as secondary to other ambitions. We share aspiring terraformers' wonder of Mars. Enabling a human presence there would be an outstanding feat of technical ingenuity and a testament to the capabilities of collaborative institutions that will inspire generations around the globe to come. But we should pursue that ambition only if we can do so in a manner that prioritizes long-term stewardship over the pressures of expansionism, before irreplaceable scientific evidence is lost forever.

Standing amid the imagined ruins of the Library of Alexandria in the television series *Cosmos*, Sagan warned of those who, through fear, ignorance, or the pursuit of power, destroy treasures that belong to all of us. He issued a charge that now extends to a world hundreds of millions of kilometers away: “We must not let it happen again.”

**Acknowledgments**

The authors are grateful to Christopher Edwards for facilitating the initial discussions for this article. Work on this article was funded in part by the Dutch Research Council (NWO project PGW.23.006). This article does not represent the views of Dartmouth College, Utrecht University, Northern Arizona University, Laboratoire de Météorologie Dynamique, or Laboratoire d’Astrophysique de Bordeaux. The views expressed are solely those of the authors.

**Author Information**

Ari Koeppel (ari.koeppel@Dartmouth.edu), Dartmouth College, Hanover, NH; Lonneke Roelofs and Santiago Vrech, Utrecht University, Netherlands; Alicia M. Rutledge, Northern Arizona University, Flagstaff; Martin Turbet, Laboratoire de Météorologie Dynamique, Sorbonne Université and Laboratoire d’Astrophysique de Bordeaux, Université de Bordeaux, Pessac, France.

**Main Image:**

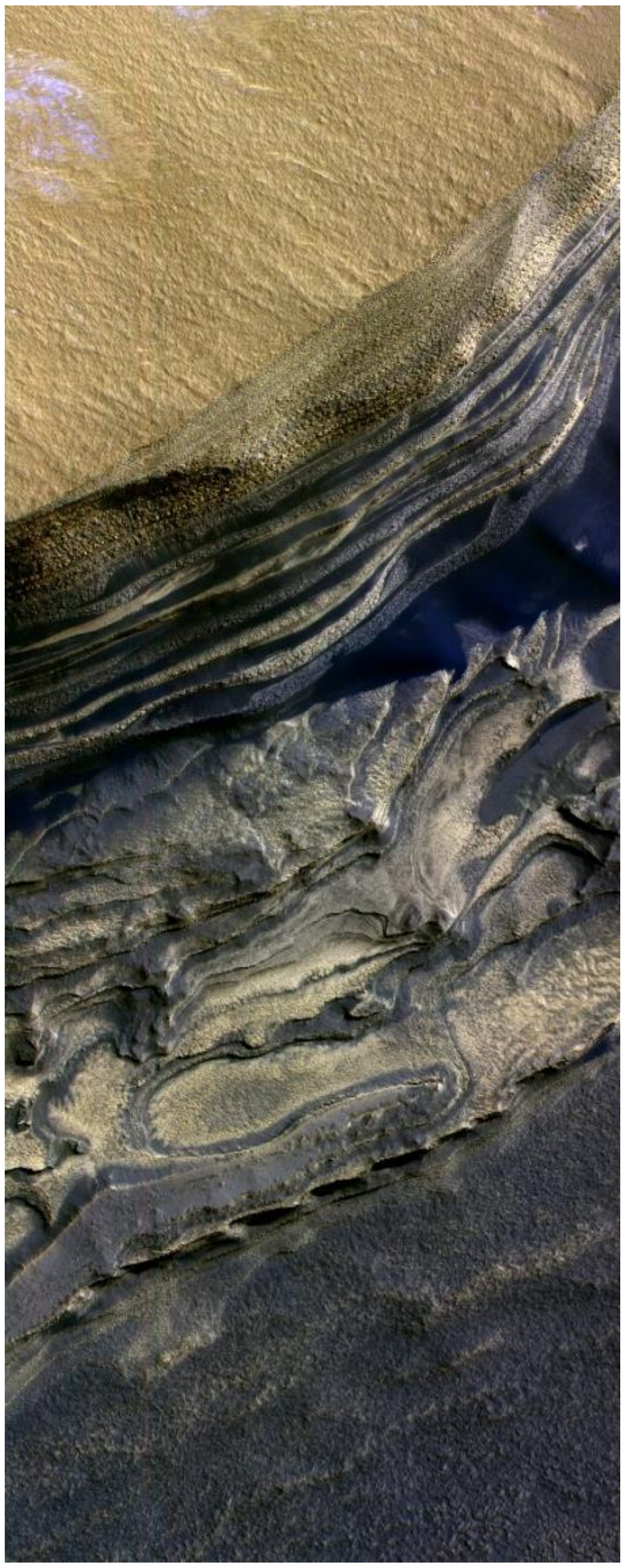

**Caption/Credit:** Layers of ice on Mars, like those viewed in Chasma Boreale by the High Resolution Imaging Science Experiment (HiRISE), hold a fragile record of climate fluctuations. Credit: NASA/JPL-Caltech/Univ. of Arizona https://science.nasa.gov/photojournal/false-color-image-of-north-polar-layered-deposits-in-head-scarp-of-chasma-boreale/